\documentclass[reprint,showpacs,prb,superscriptaddress]{revtex4-1}
\pdfoutput=1

\usepackage{amsmath}
\usepackage{epsf}
\usepackage{graphicx}
\usepackage{dcolumn}
\usepackage{bm}

\renewcommand{\k}{\mathbf{k}}

\newcommand{\beq}{\begin{equation}}
\newcommand{\eeq}{\end{equation}}
\newcommand{\be}{\begin{equation}}
\newcommand{\ee}{\end{equation}}
\newcommand{\beqa}{\begin{eqnarray}}
\newcommand{\eeqa}{\end{eqnarray}}
\newcommand{\bea}{\begin{align}}
\newcommand{\eea}{\end{align}}
\usepackage{amssymb}
\usepackage{array}
\usepackage{amsmath}
\usepackage{graphicx}\usepackage{graphics}
\usepackage{dcolumn}
\usepackage{bm}\usepackage{varioref}

\let\bm\undefined
\newcommand{\bm}[1]{\mathbf{#1}}

\newcommand{\pd}{\partial}

\newcommand{\hbk}{\bm{k}}

\usepackage{varioref}
\usepackage{wrapfig}
\usepackage{etoolbox}
\usepackage{color}
\AtEndEnvironment{thebibliography}{
}
\begin{document}

\def\boldsymbol#1{\mbox{\boldmath$#1$}}

\title{Speckle statistics of entangled photons}

\author{Avraham Klein}
\affiliation{The Racah Institute of Physics, The Hebrew University of Jerusalem, 91904, Israel}
\author{Oded Agam}
\affiliation{The Racah Institute of Physics, The Hebrew University of Jerusalem, 91904, Israel}
\author{Boris Spivak}
\affiliation{Physics Department, University of Washington, Seattle, WA, USA}

\date{\today}
\begin{abstract}
  We consider the propagation of several entangled photons through an elastically scattering medium and study statistical properties of their speckle patterns. We find the spatial correlations of multi-photon speckles and their sensitivity to changes of system parameters. Our analysis covers both the directed-wave regime, where rays propagate almost ballistically while experiencing small-angle diffusion, and the real-space diffusive regime. We demonstrate that long-range correlations of the speckle patterns dominate experimental signatures for large-aperture photon detectors. We also show that speckle sensitivity depends strongly on the number of photons $N$ in the incoming beam, increasing as $\sqrt{N}$ in the directed-wave regime and as $N$ in the diffusive regime.
\end{abstract}

\maketitle
\section{introduction}

The statistics of the speckle pattern of classical coherent light propagating through a disordered medium has been the focus of considerable research
\cite{Zyuzin1987,Feng1988,Spivak1991}. This system can be described by a scalar field $\psi(\bm x)$ satisfying the Helmholtz equation
\be
\label{eq:helmholtz}
\bar{k}^2 n^2(\bm x) \psi(\bm x)+ \nabla^2 \psi(\bm x)=0,
\ee
Here, $\bar{k}$ is the wave number, and $n=n_0+\delta n(\bm x)$ is the index of refraction, where $n_0$ and $\delta n$ denote, respectively,  the average index of refraction and its spatial fluctuation.  (In this paper we ignore the vector nature of the photons. Also, without loss of generality we set $n_0 = 1$.) Customarily one assumes that  $\delta n(\bm x)$ is a gaussian random field with zero mean and a correlation function,
\be
\langle \delta n(\bm x)\delta n(\bm x')\rangle= g(\bm x-\bm x').
\label{eq:g}
\ee
Here angular brackets denote ensemble averaging over the random realizations of the disorder.  Within the Born approximation the scattering probability per unit length is determined by the Fourier transform of the disorder correlation function: 
\be
{\cal G}(\bm s)= \frac{\bar{k}^4}{\pi}\int d^3 r g(\bm x) \exp(i\bar{k}{\bf s}\cdot\bm x) \label{eq:CalG}
\ee
where $\bm s$ is a unit vector associated with the change in the ray direction. Thus, the elastic mean free path $\ell$ and the transport mean free path $\ell_{tr}$ are given by:
\beqa
\ell^{-1}&=& \int d^2s' {\cal G}({\bf{s}}-{\bf{s}}') \\
\ell^{-1}_{tr}&=&\int d^2s' {\cal G}(\bf{s}-\bf{s}')(1-\bf{s}\cdot\bf{s}')
\eeqa
where the integration of over different directions is normalized such that $\int d^2s=1$. 

In many systems it is possible to identify two distinct regimes: when the transport mean free path is much longer than the mean free path, and the system size $L$ satisfies the condition, $\ell \ll L \ll \ell_{tr}$, then photons experience a series of small-angle scattering events. In this  ``directed-wave'' regime, as it is called, waves propagate almost ballistically through the sample but experience small diffusive changes in direction. The second regime is when the sample size is much bigger than the transport mean free path, $L \gg \ell_{tr}$, so that the photons diffuse through the sample.  

\begin{figure*}
\includegraphics[width=\textwidth]{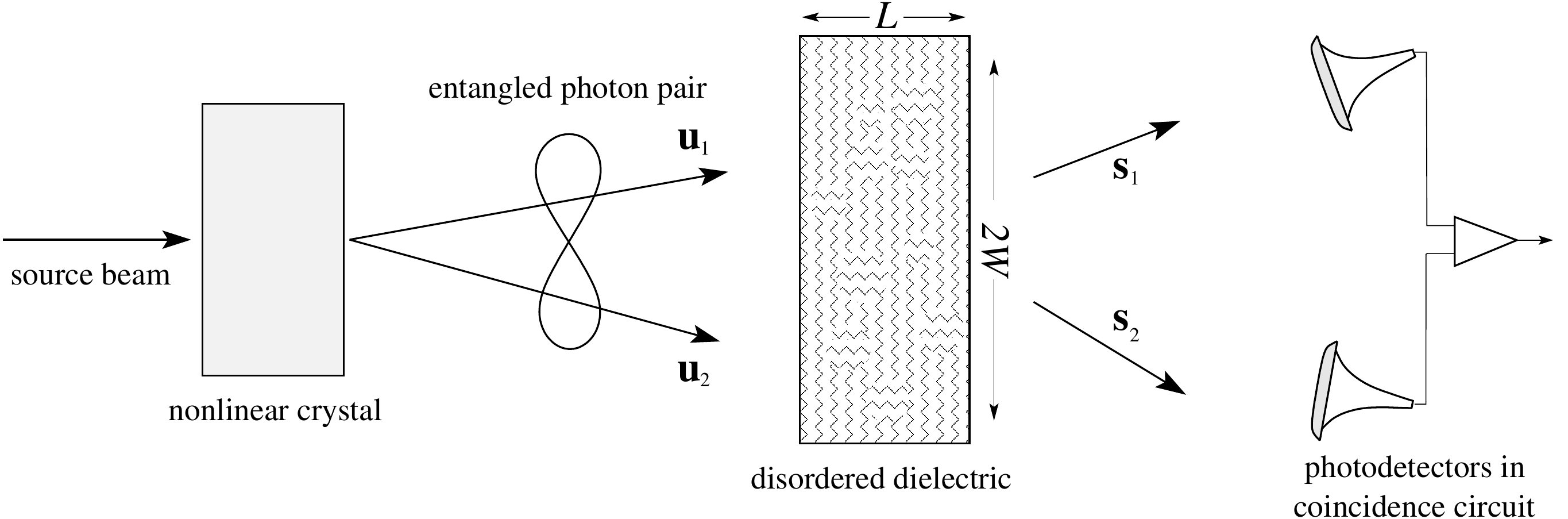}
\caption{Illustration of an experimental system for measuring a biphoton speckle pattern. The entangled photons are generated by spontaneous down-conversion in a nonlinear crystal. The two photons pass through a scattering medium and are collected by two detectors in a coincidence circuit.}
\label{fig0}
\end{figure*}

Recent advances\cite{Peeters2010,DiLorenzoPires2012,Lee2006,Cherroret2011} have allowed experimentalists to entangle several photons, see for example Ref. \onlinecite{Yao2012}, creating an excellent setup for studying multi-photon speckle statistics. The experimental system for measuring speckle patterns of entangled photons consists of a multi-photon source, emitting a beam that passes through a disordered and elastically scattering medium. The photons are then collected by a set of photodetectors in a coincidence circuit. Fig. \ref{fig0} shows a sketch of such a system, where for clarity we show a two-detector setup. If the detectors have an angular aperture which is much smaller than the typical size of a classical speckle, they measure in essence the biphoton current $I_2(\bm s,\bm s')$ going in directions $\bm s$ and $\bm s'$.

In a disordered system $I_2$ is a random quantity. Its statistics were studied by Beenakker, Vanderbos and van Exter \cite{Beenakker2009}, using a random matrix theory (RMT) approach. In particular they showed that the fluctuations in the biphoton current (i.e. the number of coincident detections per unit time by two photon detectors) $\delta I_2(\bm s,\bm s')=I_2(\bm s,\bm s')-\langle I_2(\bm s,\bm s') \rangle$ satisfy the relation
\begin{equation}
\frac{\langle \delta I_2^2(\bm s,\bm s')\rangle}{\langle I_2(\bm s,\bm s')\rangle^2} = \mbox{Tr} \rho^2 + 2  \mbox{Tr} \left( \rho^{(1)} \right)^2 
\label{eq:beenakker-main}
\end{equation}
Here $\rho$ is the two photon density-matrix, while $\rho^{(1)}$ is the reduced density-matrix obtained by tracing over the states of one of the photons. Eq.~(\ref{eq:beenakker-main}) shows that intensity fluctuations of the biphoton speckle patterns encode the information about the purity of system as well as its degree of entanglement.

However, RMT cannot account for spatial correlations which exist in the speckle patterns. In this paper, we study these correlations, show they are long-ranged, and that they become important when detectors collect photons from a large enough solid angle. These long range correlations depend only on some reduced density matrix, and therefore they do not contain information about the purity of the system. We also study the speckle pattern's sensitivity to a change in parameters such as the photon wavepacket's incidence angle or frequency, and to changes in the scattering potential. The description of this sensitivity is also beyond RMT.

In this work we consider a slab geometry for the scattering medium. The slab cross-sectional area is $\mathcal A \sim W^2$  and its thickness is $L$.  Our main concern is to describe the statistical properties of the current associated with the scattered beam of entangled photons, given by the following expression:
\begin{widetext}
\be
\label{eq:I_N-def}
I_N(\bm s_1,\bm s_2, \cdots \bm s_N)=\beta_N \sum_{\tiny \begin{array}{c} u_1\cdots u_N \\ \tilde{u}_1 \cdots \tilde{u}_N \end{array} }\rho_{u_1\cdots u_N;\tilde{u}_1 \cdots \tilde{u}_N} \prod_{j=1}^{N}  \psi_{\bm{u}_j}({\bm s}_j) \psi^*_{\tilde{\bm{u}}_j}({\bm s}_j) 
\ee
\end{widetext} 
Here $\beta_N= c^N (\Delta t)^{N-1}$, where $c$ is the speed of light and  $\Delta t$ is the resolving time of the coincidence circuit, while  $\rho_{u_1\cdots u_N;\tilde{u}_1 \cdots \tilde{u}_N}$ is the density matrix of the incoming beam of $N$ entangled photons. Finally, $\psi_{\bm{u}}(\bm s)$ is the component of the scattered classical wave function in direction $\bm s$ associated with an incoming plane wave with direction $\bm u$. In order to simplify our final expressions, we assume that the incoming wavepacket can be decomposed into well-separated plane-waves, $|\bm u_i - \bm u_j| > 1 /(\bar{k}L\theta_L)$ in the directed-wave regime and $|\bm u_i - \bm u_j| > 1/(\bar{k}L)$ in the diffusive regime. Here $\theta_L$ is the typical angular spread of the outgoing beam after crossing the sample (we will give an explicit definition for it later, see Eq. \eqref{AvI2f}).

\section{Multiphoton speckle patterns in the directed wave regime}
\label{sec:mult-speckle-patt}
We begin our discussion with the directed wave regime $\ell \ll L \ll \ell_{tr}$. To simplify the discussion, in what follows we shall consider the case of a biphoton current, $N=2$. The results will be generalized to $N$-photons in Sec. \ref{sec:gener-n-entangl}.

In the directed waves regime it will be convenient to decompose  the three dimensional vectors into components that are parallel and transverse to the ray propagation direction which we choose  be the $z$ direction, thus
\begin{equation}
  \label{eq:vector-dec}
  \bm x = (\bm r, z),~~~~~~~ \bm s = \frac{(\bm k,1)}{\sqrt{1+ k^2}} \simeq( \bm k, 1),
\end{equation}
where ${\bf r}$ and ${\bf k}$ are two dimensional vectors , and in the directed waves regime $k^2 \ll 1$.

\begin{figure*}
  \begin{minipage}{0.45\textwidth}
    \includegraphics[width=0.4\textwidth,clip=true,trim=10 2 4 2, angle=-90]{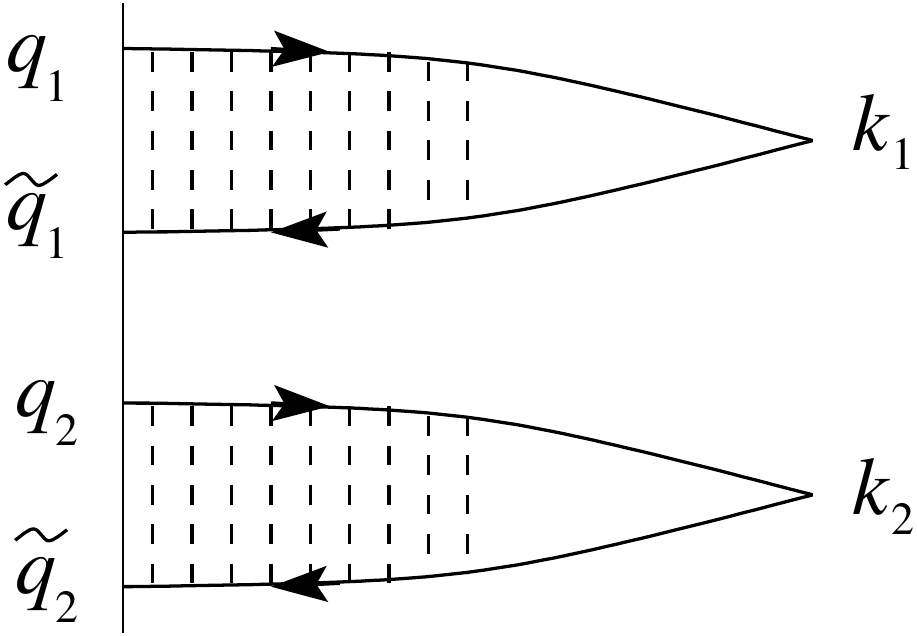}
    \caption{The leading order diagrams (in $1/\bar{k}\ell$) describing the average biphoton current, $\langle I_2(\hbk_1,\hbk_2) \rangle$. Solid lines represent the average Green's functions while dashed lines represent the disorder.}
    \label{fig2}
  \end{minipage}
  \begin{minipage}{0.45\textwidth}
    \includegraphics[width=0.3\columnwidth, angle=-90]{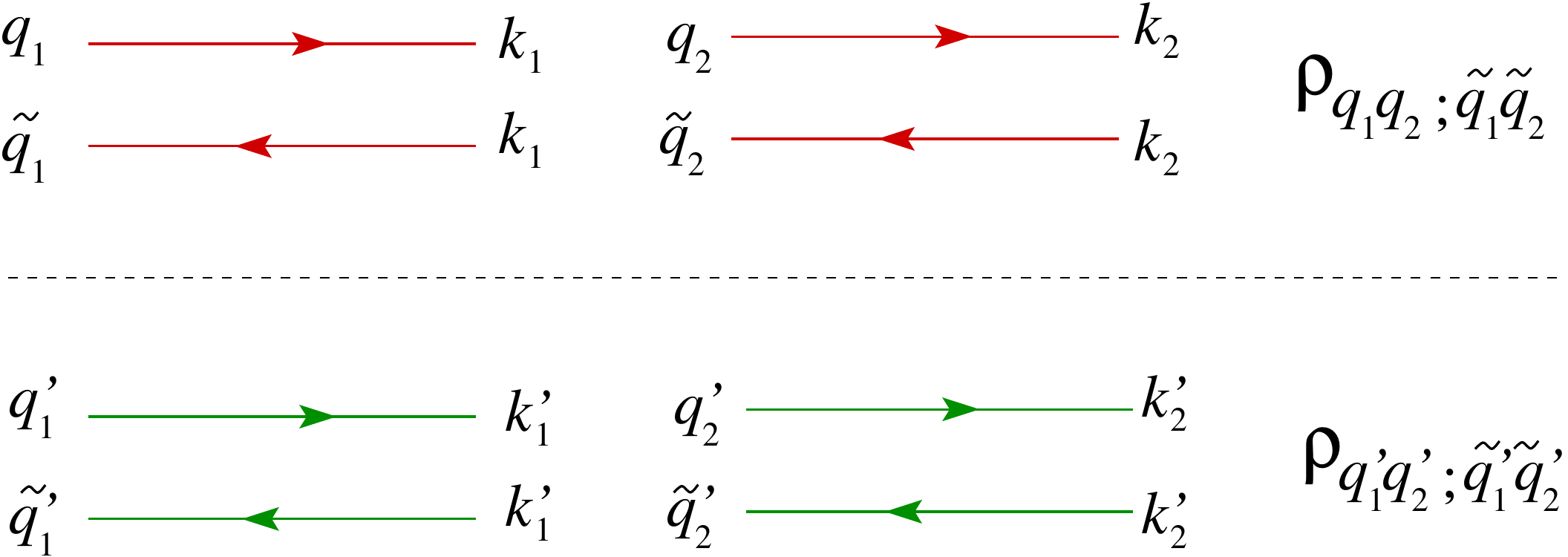}
    \caption{(Color online) Green's functions associated with the correlation function of the biphoton current before disorder averaging. In order to obtain the connected part of the correlation function, one must pair at least one of the Green's function that are above the dashed line (red) with one below the dashed line (green).}
    \label{fig3}
  \end{minipage}\vspace{1cm}
  \begin{minipage}{0.45\textwidth}
    \vspace{0cm}
    \includegraphics[width=0.4\columnwidth, angle=-90]{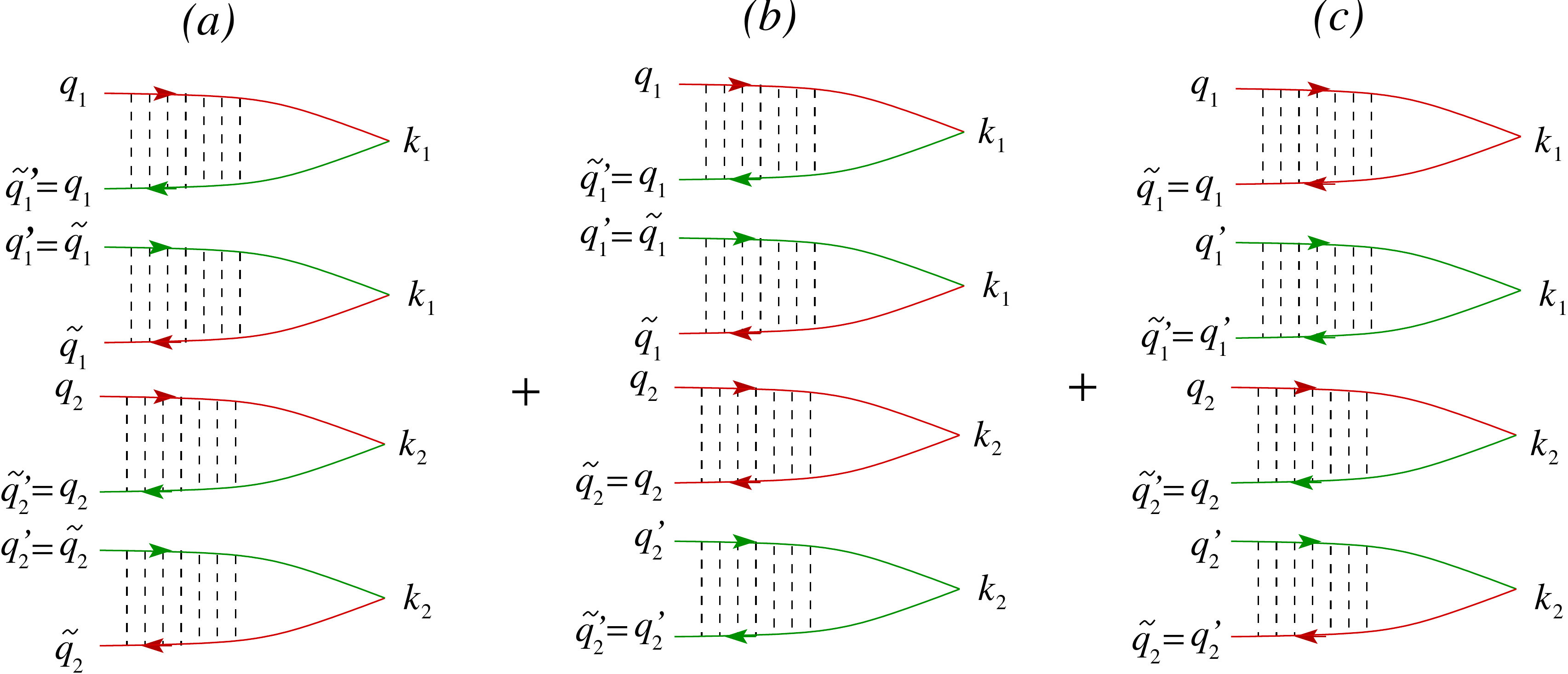}
    \caption{(Color online) The leading-order diagrams associated with the correlation function $\mathcal{K}(\hbk_1,\hbk_2;\hbk_1,\hbk_2) =   \langle \delta I_2^2(\hbk_1,\hbk_2)\rangle$, corresponding to an experimental setup where both detectors are held fixed at given angles $\hbk_1,\hbk_2$. These diagrams yield the RMT result in Eq. \eqref{eq:beenakker-main}.}
    \vspace{1cm}
    \label{fig4}
  \end{minipage}
  \begin{minipage}{0.45\textwidth}
    \includegraphics[width=0.5\columnwidth, angle=-90]{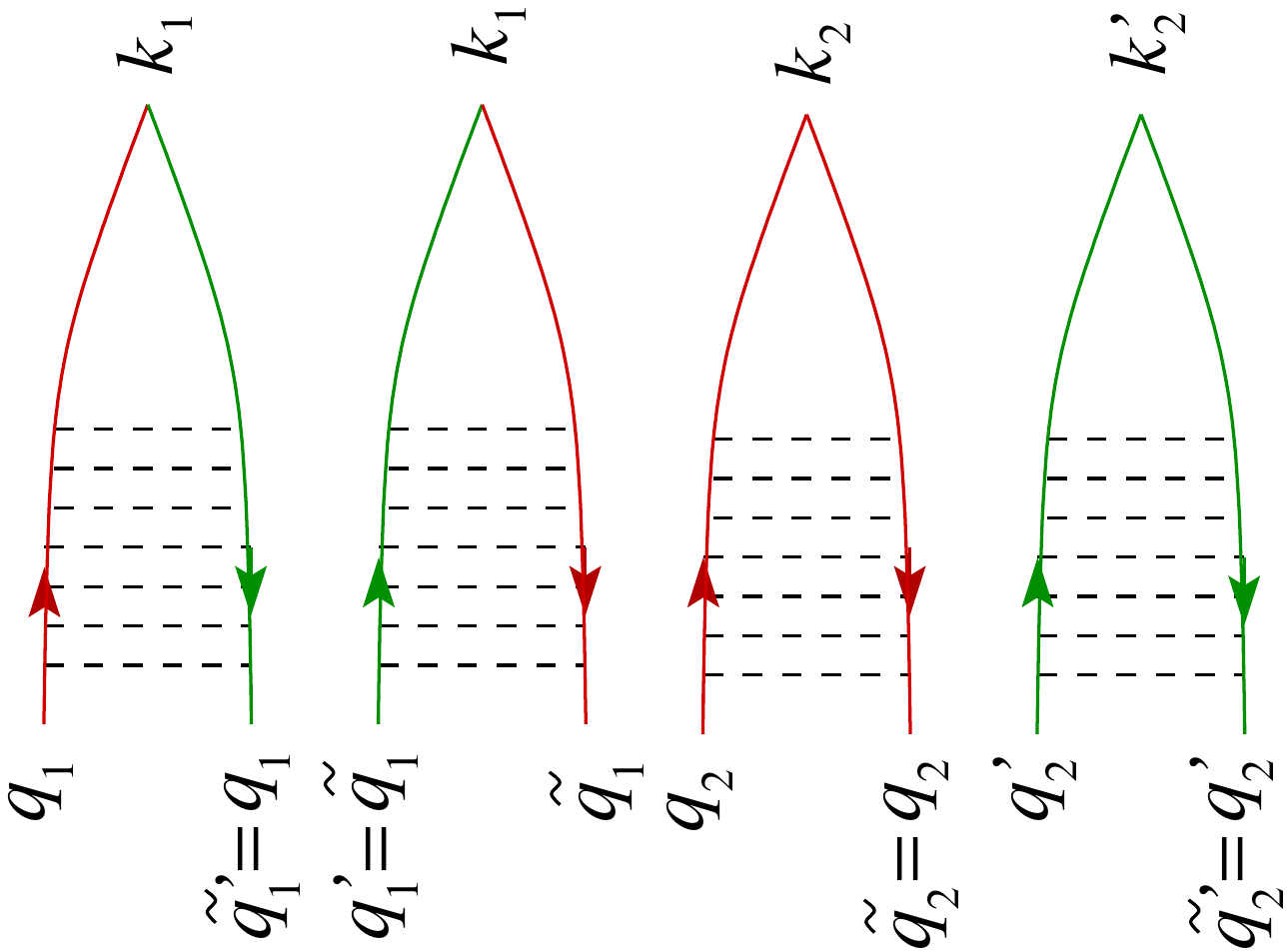}
    \caption{The leading-order diagram associated with the  correlation function at three different angles $\mathcal{K}(\bm{k}_1,\bm{k}_2;\bm{k}_1,\bm{k}'_2)$, corresponding to one fixed detector and one detector that changes position.}
    \label{fig5}
  \end{minipage}
\end{figure*}

\subsection{The disorder averaged biphoton current}

The average biphoton current can be written as
\begin{widetext}
\beqa
\langle I_2(\bm k_1,\bm k_2)\rangle &=& \beta_2 |A_0|^4 \sum_{q_1,q_2, \tilde{q}_1\tilde{q}_2}
\rho_{q_1q_2;\tilde{q}_1 \tilde{q}_2}
\int \exp\left\{ i \bar{k}\left[ {\bf k}_1\cdot({\bf r}_1'-\tilde{{\bf r}}_1')+ {\bf q}_1\cdot{\bf r}_1-\tilde{\bf q}_1\cdot\tilde{{\bf r}_1}+ {\bf k}_2\cdot({\bf r}_2'-\tilde{{\bf r}}_2')+ {\bf q}_2\cdot {\bf r}_2-\tilde{{\bf q}}_2 \cdot\tilde{{\bf r}}_2)\right]\right\}
\nonumber
\\ &\times&
\left\langle G({\bf r}_1,0; {\bf r}_1', L) G^*(\tilde{{\bf r}}_1,L; \tilde{{\bf r}}_1', 0)G({\bf r}_2,0; {\bf r}_2', L) G^*(\tilde{{\bf r}}_2,L; \tilde{{\bf r}}_2', 0) \right\rangle, \label{AvI2}
\eeqa
\end{widetext}
where integration is taken over all space variables. Here $G({\bf r},z; {\bf r}', z')$ is the Green's function of (the classical) Eq. \eqref{eq:helmholtz}. 
In the directed-wave limit, the paraxial approximation applies and the axis of propagation direction ($z$ in our notation) plays a role similar to time in the propagation of the wave function $\psi ({\bf r},z)$. Therefore in this regime $\psi_{\bm u}(\bm s)$ from Eq. \eqref{eq:I_N-def} takes the form:
 \be
\psi_{{\bf q}} ({\bf k})= A_0 \int  d^2r d^2r' G({\bf r},L; {\bf r}', 0) e^{i\bar{k}\left({\bf q}\cdot {\bf r}'- {\bf k}\cdot {\bf r} \right)}
\ee
where $A_0$ is the incoming wave amplitude, into which
we also absorb a phase which plays no role in the final results.

Since the Green's function $G({\bf r},z; {\bf r}', z')$ is a classical object, we can apply the same techniques developed in the context of propagation of classical waves in disordered systems to the task of finding the average and correlation function of $I_N$. To leading order in $1/ {\bar{k} \ell}$ it is  given by the diagram shown in Fig. \ref{fig2}, representing diffusion in angular space. 

A description of the diagrammatic technique for classical waves in the directed-wave regime can be found in Ref. \onlinecite{Agam2007}. Here we bring only final results, but a short review of the technique and some details of our calculations can be found in Appendix \ref{sec:app-two-point-corr}. In the directed-wave regime we find that Eq. \eqref{AvI2} reduces to:
\begin{widetext}
\beqa
\langle I_2({\bf k}_1,{\bf k}_2)\rangle  \simeq  \beta_2 |A_0|^4\mathcal{A}^2 \sum_{q_1,q_2}\rho_{q_1q_2;q_1 q_2}
\frac{ 1}{\left( 2\pi \theta_L^2\right)^2}
\exp\left[ -\frac{ ({\bf k}_1- {\bf q}_1)^2+({\bf k}_2- {\bf q}_2)^2 }{ 2\theta_L^2}\right]
\label{AvI2f}
\eeqa
\end{widetext}
where $\mathcal{A}$ is the cross-sectional area of the slab, $\theta_L^2=2 D_\theta L$ is the angular spread of the outgoing beam due to scattering within the slab, and $D_\theta = 1/2\ell_{tr}$ is the diffusion coefficient in angular space. We have also, in the interest of brevity, rewritten $\rho$ in terms of the two dimensional $q_j$. From now on we shall consider the limit $|{\bf q}_i-{\bf k}_j| \ll \theta_L$, for all $i$ and $j$, namely that all detectors are positioned within the angular spread of the outgoing beam. In this limit Eq. \eqref{AvI2f} reduces to
\beqa
\langle I_2({\bf k}_1,{\bf k}_2)\rangle \equiv  \langle I_2 \rangle \simeq  \beta_2 \left(\frac{|A_0|^2\mathcal{A}}{ 2\pi \theta_L^2}\right)^2
\label{AvI2ff}.
\eeqa
To obtain this formula, we have used the normalization condition: $\mbox{Tr} \rho = \sum_{q_1,q_2} \rho_{q_1 q_2; q_1 q_2}=1$.

\subsection{The 2-photon current correlation function (small aperture detectors) }

Let us now calculate the biphoton current correlation function:
\begin{equation}
\label{eq:2-photon-corr}
\mathcal{K}(\bm k_1,\bm k_2;\bm k'_1,\bm k'_2) =   \langle \delta I_2(\bm k_1,\bm k_2)\delta I_2(\bm k'_1,\bm k'_2)\rangle.
\end{equation}
Before disorder averaging, the diagrammatic representation of $I_2(\bm k_1,\bm k_2)I_2(\bm k_1',\bm k_2')$ is as shown in Fig. \ref{fig3}. After disorder averaging, the biphoton current correlator Eq. \eqref{eq:2-photon-corr} is represented by the diagrams in Figs. \ref{fig4},\ref{fig5} and \ref{fig6}. We now discuss three different limits for the correlation function which correspond to three experimental setups with two small aperture detectors:  The first is when both  detectors are held fixed at two angles; the second is when one detectors is held fixed and  the position of the other is changed; the third is when the positions of both detectors are changed. In all cases we shall assume that the detectors are separated by an angle much larger than 
\be
\theta_W=\frac{1}{\bar{k}W},
\ee
 where $W$ is the slab width (see Fig. \ref{fig0}). It will be shown below that $\theta_W$ is the typical angular size of a speckle. 
  
\subsubsection{The correlation function at two different directions $\mathcal{K}(\bm k_1,\bm k_2;\bm k_1,\bm k_2) =   \langle \delta I_2^2(\bm k_1,\bm k_2)\rangle$ (the RMT limit)}

The case when both detectors are held fixed is the one described in Ref. \onlinecite{Beenakker2009} by RMT. It is given by choosing $\bm k_1 = \bm k_1', \bm k_2 = \bm k_2'$. The leading contribution in this case  is described by the diagrams shown in Fig. \ref{fig4}, which yield
\begin{widetext}
\beqa
 \mathcal{K}(\bm k_1,\bm k_2;\bm k_1,\bm k_2) =
\langle I_2\rangle^2 \sum_{q_1,q_2,\tilde{q}_1,\tilde{q}_2} \left[\rho_{q_1q_2;\tilde{q}_1\tilde{q}_2} \rho_{\tilde{q}_1\tilde{q}_2;q_1q_2}+   2\rho_{q_1q_2;\tilde{q}_1q_2} \rho_{\tilde{q}_1\tilde{q}_2;q_1\tilde{q}_2}\right]
\label{RMTK}
\eeqa
\end{widetext}
The first contribution is associated with the diagram shown in Fig. \ref{fig4}a, while the second comes from Figs. \ref{fig4}b,\ref{fig4}c. Performing the summation yields the RMT result, Eq. \eqref{eq:beenakker-main}, after noting that the reduced density matrix $\rho^{(1)}_{q_1;\tilde{q}_1} = \sum_{q_2} \rho_{q_1q_2;\tilde{q}_1q_2}$. 

\subsubsection{ The correlation function at three different directions $\mathcal{K}({\bf k}_1,{\bf k}_2;{\bf k}_1,{\bf k}'_2)$}

The case when one detector is held fixed while the other changes position is described by choosing $\bm k_1 = \bm k_1'$, $|\bm k_2 - \bm k_2'| \gg \theta_W)$. In this case the leading order diagram in $1/(\overline{k}\ell)$, which is shown in Fig. \ref{fig5}, yields:
\beqa
 \mathcal{K}({\bf k}_1,{\bf k}_2;{\bf k}_1,{\bf k}'_2) =
\langle I_2\rangle^2    \mbox{Tr} \left(\rho^{(1)}\right)^2. 
\label{K1}
\eeqa

\subsubsection{\it The correlation at four different directions: $\mathcal{K}({\bf k}_1,{\bf k}_2;{\bf k}'_1,{\bf k}'_2)$}
\label{sec:it-correlation-at}
Consider now the third case when the positions of both detectors are changed. The leading order diagrams in this case are shown in Fig. \ref{fig6}. The result is:

\begin{figure*}
\includegraphics[width=0.5\textwidth, angle=-90]{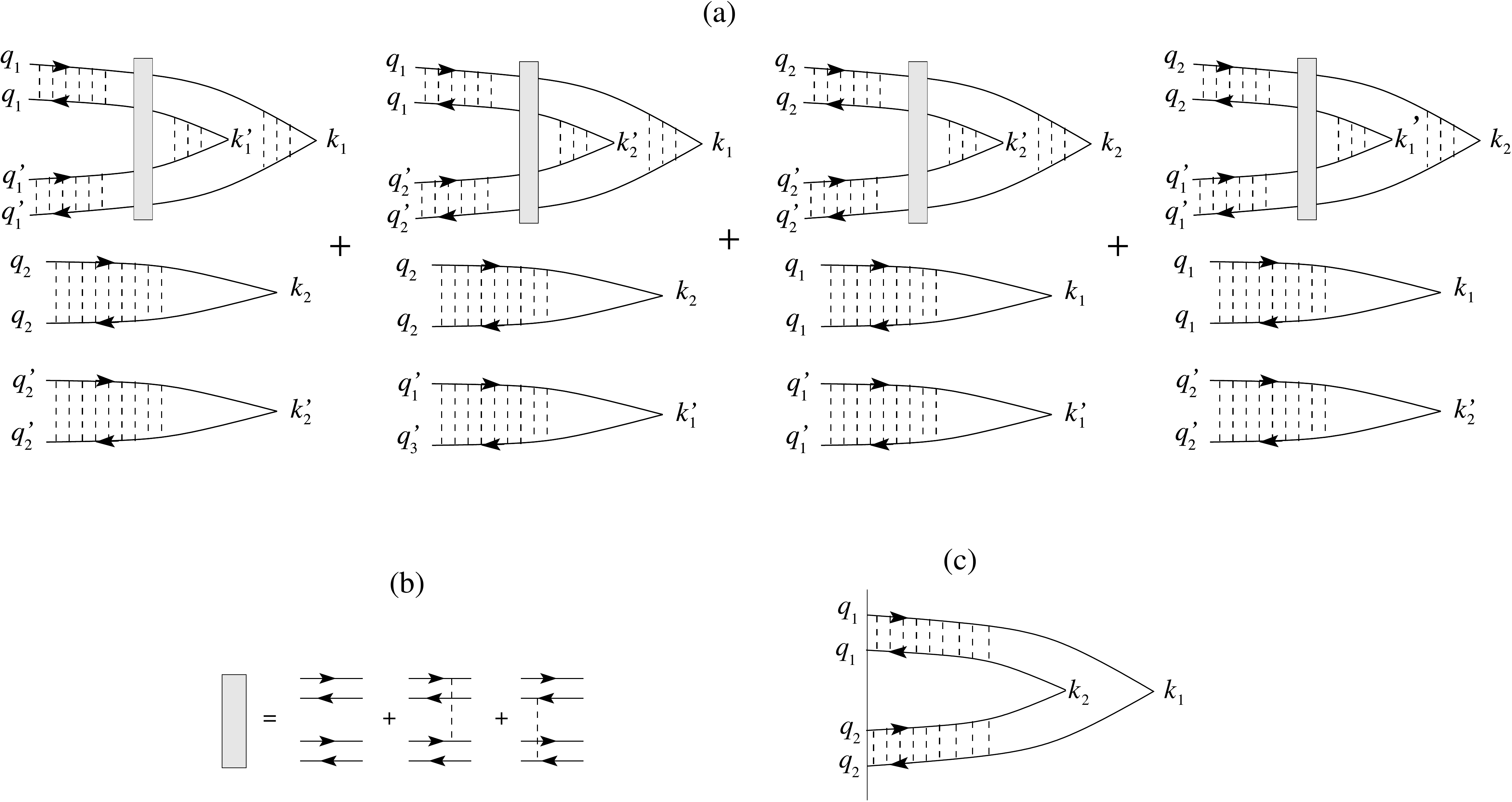}
\caption{(a) The leading order diagrams contributing to the correlation function at four different angles $\mathcal{K}(\bm{k}_1,\bm{k}_2;\bm{k}'_1,\bm{k}'_2)$. (b) Diagram describing a Hikami box. (c) Diagram describing the short range limit of the correlation function.}
\label{fig6}
\end{figure*}
\be
\label{eq:directed-K}
\mathcal{K}(\bm k_1,\bm k_2;\bm k'_1,\bm k'_2)=  \langle I_2\rangle^2\mbox{ Tr} \left(\rho^{(1)}\right)^2  \sum_{ij=1,2}\mathcal{C}({\bf k}_i,{\bf k}'_j)
\ee
where $\mathcal{C}({\bf k},{\bf k}')$ is a correlation function which is calculated in Appendix A. In the regime of interest,  $\theta_L \gg k_1, k_2, q_1, q_2$, this correlation function reduces to a function of the difference of the directions $\mathcal{C}({\bf k}_1,{\bf k}_2 ) \to \mathcal{C}_0(|{\bf k}_1-{\bf k}_2|)$ where
\begin{subequations}\label{eq:C0-direct-all}
\be
\label{C0}
\mathcal{C}_0(k) \simeq-\frac{1}{ \bar{k}^2 \mathcal{A}} \left\{ \begin{array}{cc}
2\pi \ell {\cal G}(k) & \theta_W \ll k \ll \theta_0 \\ \frac{1}{2\theta_L^2} & \theta_0\ll k \ll \theta_L 
\end{array} \right. .
\ee
Here $\theta_0^2 = \ell/\ell_{tr}$ is the typical scattering angle at distances on order of the elastic mean free path, ${\cal G}$ was defined in (\ref{eq:CalG}), and we assume the disorder to be isotropic. (Note that the Born approximation is valid when $\theta_0 \gg 1/(\bar{k}\ell)$.) 

For small scattering angles, $k \ll \theta_0$, the correlation function can be calculated from diagram (c) in Fig. \ref{fig6}. The result is:
\beqa
\mathcal{C}_{0}(k) = \left(\frac{2 J_1(k/\theta_W)}{k/\theta_W}\right)^2. \label{C-Short}
\eeqa
\end{subequations}
Here $J_1(x)$ is a Bessel function of the first kind and we have assumed a slab geometry with a circular cross section of radius $W$.

\begin{figure}
\includegraphics[width=\columnwidth,clip=true,trim=5 290 0 0]{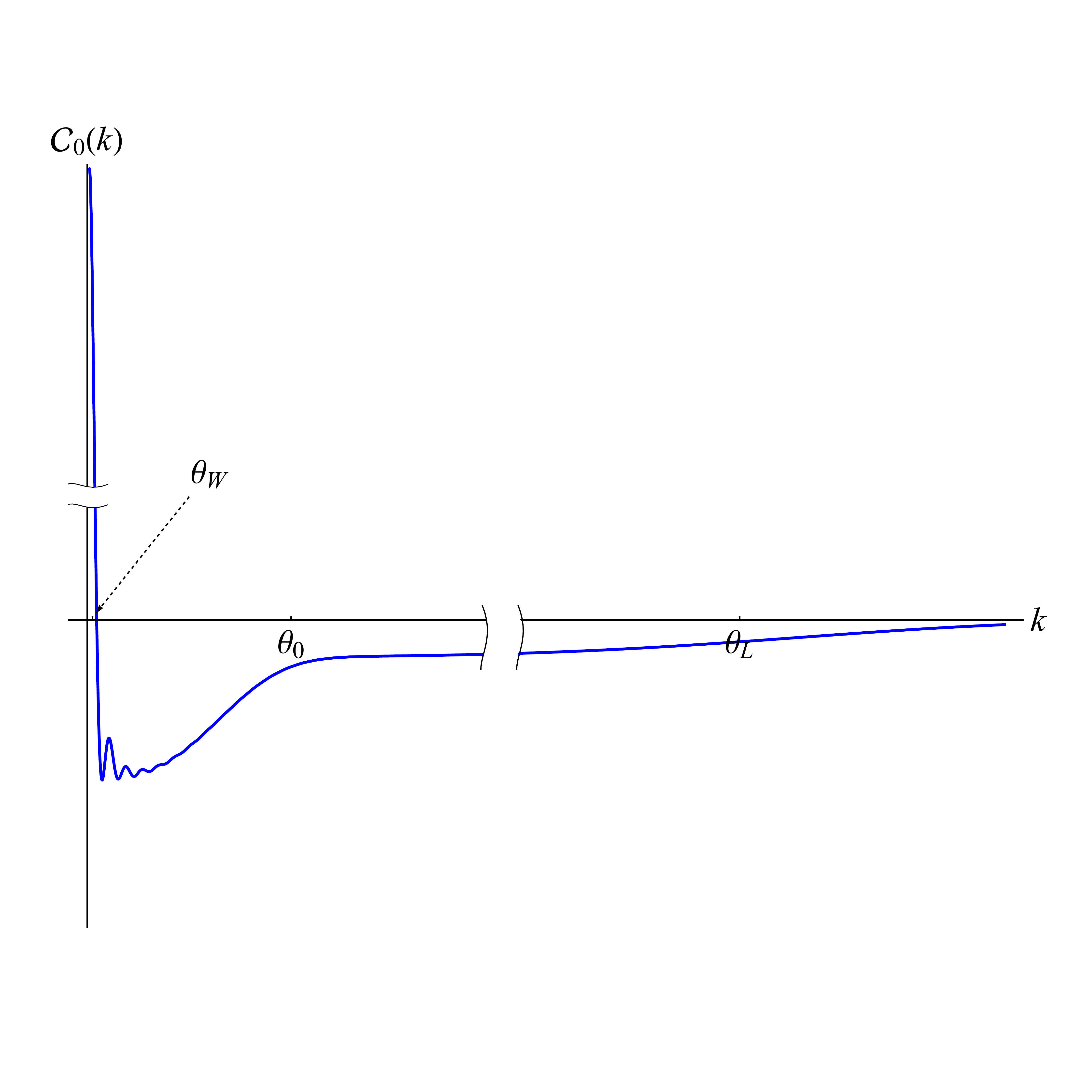}
\caption{A schematic illustration of the behavior of the correlation function $\mathcal{C}_0(k)$ whose asymptotic behavior is given by Eqs. \eqref{eq:C0-direct-all}.}
\label{fig7}
\end{figure}

 A schematic plot of ${\cal C}_0(\hat k)$ is depicted in Fig. \ref{fig7}. It is characterized by four distinct regions: At angles smaller than $\theta_W$ the correlation decays rapidly from its maximal value, 1, and it changes sign at an angle of order $\theta_W$. In the second region,  $\theta_W < k < \theta_0$ the correlation function is negative and of order $-1/(\bar{k}^2 {\cal A} \theta_0^2)$. Its absolute value decreases to a value of order $-1/(\bar{k}^2 {\cal A} \theta_L^2)$ at $k= \theta_0$. The correlation function remains essentially fixed at this value within the third region  $ \theta_0 \ll k \ll \theta_L$. Finally, for $k \gg \theta_L$ the correlation function decays as $\exp[ -k^2/(2\theta_L^2)]$ (this regime is not described by the limiting formula (\ref{C0})).  

\noindent
\subsection{Fluctuations of 2-photon currents measured with large aperture detectors}

When the detector aperture collects a beam from an angle $\theta_1 \gg \theta_W$, the long-range correlations described in the previous section will dominate the measurement. For such a setup, the total flux incident on the detector will be:
\be
P(\bm{k}, \bm{k}') = \int_{| k_1|, |k_2|<\theta_1}\!\!\!\!\!\!\!\!\!\!\!\! d^2  k_1 d^2  k_2 I_2(\bm{k}+ \bm{k}_1,\bm{k}'+\bm{k}_2)
\ee
and the mean square variation of $P$ is then
\be
\langle \delta P(\bm{k}, \bm{k}')^2\rangle \!=\!\!\int
 \langle \delta I_2(\bm{k}+\bm{k}_1,\bm{k}'+ \bm{k}_2) \delta I_2(\bm{k}+\bm{k}'_1,\bm{k}'+ \bm{k}'_2) \rangle \label{eq:P}
\ee
where integration is over the space, $k_1,k_2, k_1', k_2' < \theta_1$.  There are several contributions to this integral. The first comes from the short range correlation associated with RMT. The typical angular scale of these short range correlations is $\theta_W$, see Eq.~(\ref{C-Short}) and Fig. \ref{fig7}. Therefore the corresponding contribution is of order $ \mathcal{K}(\hbk,\hbk';\hbk,\hbk')\theta_1^4 \theta_W^4$. There are two additional long-ranged contributions. One  comes from the integral (\ref{eq:P}) along the lines $k_1=k_1'$ or $k_2=k_2'$ and is of order $\mathcal{K}(\hbk,\hbk';\hbk,\tilde{\hbk}') \theta_1^6\theta_W^2$, see Eq. \eqref{K1}. The other is associated with the long-range correlation shown in Eq. \eqref{eq:directed-K} and is of order  $\mathcal{K}(\hbk,\hbk';\hbk,\tilde{\hbk}') \theta_1^8 \theta_W^2/\theta_0^2$. However, this term is much smaller since $\theta_0 \gg \theta_W$. Thus assuming Tr$\rho^2$ and Tr$\left(\rho^{(1)}\right)^2$ to be of the same order, then for large enough apertures such that $\theta_1 > \theta_W$, the long-ranged correlations will dominate the signal. In addition it is clear that once the aperture is large enough, $\theta_1 \sim \theta_L$, current conservation implies that fluctuations in $P(\k,\k')$ vanish.

\subsection{Generalization to $N$ entangled photons (small aperture detectors)}
\label{sec:gener-n-entangl}
Our previous results can be readily generalized to the case of $N$ entangled photons. We continue to assume that the absolute value of all photon wave numbers is $\bar{k}$. The correlation function for the case where $\nu$ detectors are held fixed and $N-\nu$ detectors change positions is
\begin{subequations}\label{eq:main}
\begin{widetext}
\beqa
&&\langle \delta I_N(\bm{k}_1, \cdots, \bm{k}_\nu,\bm{k}_{\nu+1}, \cdots,  \bm{k}_N) \delta I_N(\bm{k}_1, \cdots, \bm{k}_\nu,\bm{k}'_{\nu+1}, \cdots,  \bm{k}'_N) \rangle \nonumber \\
&&~~~~~~=\langle I_N \rangle^{2 }\sum_{j=0}^{\nu-1} \left( \begin{array}{c} \nu \\ j \end{array} \right) \left[\mbox{Tr} \left( \rho^{(\nu-j)}\right)^2 +(1-\delta_{\nu,N}) \mbox{Tr}\left( \rho^{(\nu-j+1)} \right)^2 \sum_{i,i'=\nu+1}^{N} \mathcal{C}(\bm{k}_i,\bm{k}_{i'})\right] \label{eq:main1}
\eeqa
where $N\geq \nu \geq 1$. Here $\rho^{(N-j)}$ is the reduced density matrix obtained by tracing out $j$ photons, and $\langle I_N \rangle= \beta_N \left(\frac{|A_0|^{2}\mathcal{A}}{ 2\pi \theta_L^2}\right)^N$ is the average $N$-photon intensity
. In the case where all detectors change position the result is:
\be
\langle \delta I_N(\bm{k}_1 \cdots  \bm{k}_N)\delta I_N(\bm{k}'_1 \cdots  \bm{k}'_N)\rangle =  \langle I_N\rangle^{2}\mbox{Tr} \left( \rho^{(1)}\right)^2  \sum_{i,j=1}^{N} \mathcal{C}(\bm{k}_i ,\bm{k}'_j) \label{eq:main2}  
\ee
\end{widetext}
\end{subequations}
Thus by measuring the fluctuations in the intensity when $\nu$ detectors are fixed at their positions while and $N-\nu$ detectors  change their positions one is able to  measure all the trace of squares of the reduced density matrix, $\rho^{(N-j)}$,  obtained when $j$ photons are traced out.

\section{Diffusive wave propagation}
\begin{figure}
  \centering
  \includegraphics[width=\columnwidth]{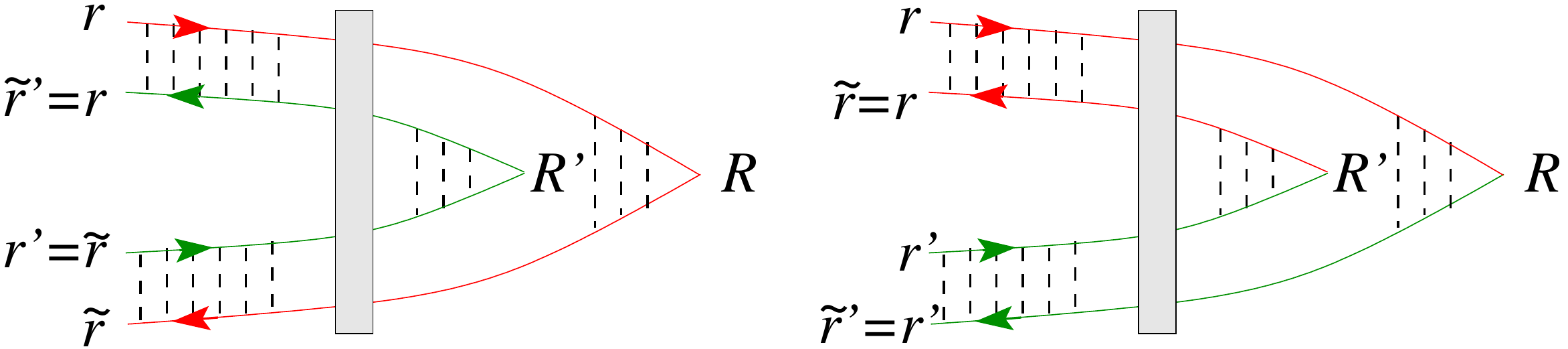}
  \caption{(Color online) The diagrams contributing to $\mathcal{K}(\bm{k}_1,\bm{k}_2;\bm{k}'_1,\bm{k}'_2)$ in the diffusive regime. The diagrams denote two types of possible interference paths and each of the the diagrams at the top of Fig. \ref{fig6}a is replaced by such a pair. In these diagrams $\bm R, \bm R'$ are conjugates to the two outgoing wavevectors, e.g. to $\hbk_1,\hbk_1'$ in the top left diagram of Fig. \ref{fig6}a.}
  \label{fig:diffusive-hikami}
\end{figure}

Consider now the case where rays diffuse in real space, i.e. the size of the system is much larger than the transport mean free path but we can neglect reflections from the slab boundaries, $\ell_{tr} \ll L \ll  W$. The diagrams describing this case are very similar to those shown for the directed wave regime. The only change is that diagrams containing Hikami boxes should be replaced by the two diagrams shown in Fig. \ref{fig:diffusive-hikami}. 
\begin{widetext}
The result for the $N$-photon correlation function is:
\begin{subequations}\label{eq:mainD}
\beqa
&&\langle \delta I_N(\bm{k}_1, \cdots, \bm{k}_\nu,\bm{k}_{\nu+1}, \cdots,  \bm{k}_N) \delta I_N(\bm{k}_1, \cdots, \bm{k}_\nu,\bm{k}'_{\nu+1}, \cdots,  \bm{k}'_N) \rangle \nonumber \\
&&=\langle I_N \rangle^{2 }\sum_{j=0}^{\nu-1} \left( \begin{array}{c} \nu \\ j \end{array} \right) \left\{\mbox{Tr} \left( \rho^{(\nu-j)}\right)^2 +(1-\delta_{\nu,N})\left[ \mbox{Tr}\left( \rho^{(\nu-j+1)} \right)^2 (N-\nu-1)^2 \tilde{\mathcal{C}}(0)+ \sum_{i,i'=\nu+1}^{N} \tilde{\mathcal{C}}(|\bm{k}_i-\bm{k}_{i}'|)\right]\right\} \label{eq:mainD1}
\eeqa
where now $ \langle I_N\rangle = \beta_N \left(\frac{|A_0|^2\mathcal{A}}{2\pi \alpha}\frac{\ell_{tr}}{L}\right)^N $ is the average $N$-photon intensity, $\alpha$ is a factor of order one depending on geometry, and\cite{Akkermans2007}

\begin{equation}
  \label{eq:C0-diffusive}
  \tilde{\mathcal{C}}(k) \simeq 
      \begin{cases}
        \left(\frac {2 W J_1( k/\theta_W)}{L \sinh(k/\varphi_L)}\right)^2 & k < \theta_W\\
        \frac{2\pi}{\overline{k}^2\mathcal{A}}\frac{L}{\ell_{tr}} & \theta_W \ll k \ll \varphi_L \\
        \frac{3\pi}{\overline{k}^2\mathcal{A}} \frac{1}{\overline{k}\ell_{tr}}\frac{1}{k} &  k \gg \varphi_L
      \end{cases}
\end{equation}
where $\varphi_L = 1/\bar{k}L$. Finally, in the case where all detectors change position the result is:
\be
\langle \delta I_N(\bm{k}_1 \cdots  \bm{k}_N)\delta I_N(\bm{k}'_1 \cdots  \bm{k}'_N)\rangle =  \langle I_N\rangle^{2}\left[\mbox{Tr} \left( \rho^{(1)}\right)^2 N^2  \tilde{\mathcal{C}}(0)+ \sum_{i,j=1}^{N} \tilde{\mathcal{C}}(|\bm{k}_i-
\bm{k}'_j|)\right] \label{eq:mainD2}  
\ee
\end{subequations}
A sketch of adaptation of the well known classical treatment\cite{Akkermans2007} yielding Eq. \eqref{eq:C0-diffusive} to the multiphoton scenario appears in Appendix \ref{sec:app-diffusive-regime}. 
\end{widetext}

Considerations similar to those discussed above imply that for two large aperture detectors, fluctuations in the detected flux are dominated by the long-range part of the correlation function when $\theta_1 > \theta_W$, just as for the directed-wave regime.

\section{Sensitivity of $N$-photon speckles to change of external parameters}

\begin{figure*}
  \centering
  \includegraphics[width=0.8\textwidth]{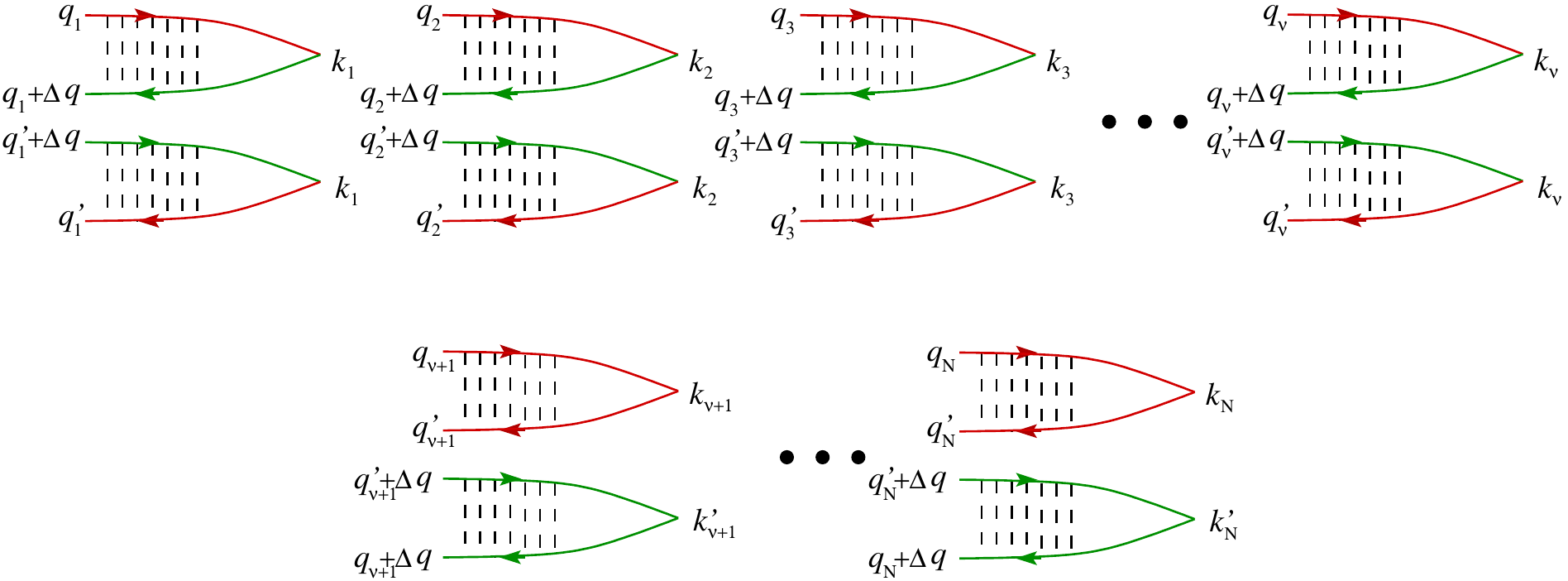}
  \caption{The leading order diagram contributing to the sensitivity correlation function $\mathcal{K}(\gamma;N)$. For simplicity we depict the diagram in for the case where the incident beam is rotated by an angle $\phi$, i.e. $\gamma = \phi = |\Delta\bm{q}|$.}
\label{fig8}
\end{figure*}

The speckle pattern of an $N$-photon beam changes as a function of external parameters such as the incidence angle, $\phi$, or frequency $\omega$, of the incoming beam, as well as realization of the scattering potential. To analyze this dependence one defines a correlation function
\be
\label{eq:sensitivity-corr}
{\cal K}(\gamma;N)=\left\langle \delta I_N^{(\gamma)} (\bm{k}) \delta I_N^{(0)} (\bm{k})\right\rangle
\ee
where $\gamma$ represents some external parameter. It is customary to characterize the sensitivity of the speckle pattern to a change $\gamma$ by a typical value $\gamma^*$, beyond which the correlation function \eqref{eq:sensitivity-corr} will have decayed significantly.

The sensitivity to changing an external parameter is described by the diagrams shown in Fig. \ref{fig8}, where red and green lines correspond to different values of $\gamma$. For the specific cases of changes in $\phi$ and $\omega$ one finds the following typical values:
\begin{align}
\phi_N^* &= 
\begin{cases} 
\frac{1}{\sqrt{N}\bar{k}L\theta_L} & \mbox{directed wave regime} \\ 
\frac{1}{N\bar{k}L} & \mbox{diffusive regime}
\end{cases}
\label{eq:sens-direct-summary}\\ 
\omega_N^* &= 
\begin{cases} 
\frac{c}{\sqrt{N}\theta_L^2 L} & \mbox{directed wave regime} \\ 
\frac{c \ell_{tr}}{N L^2} & \mbox{diffusive regime}
\end{cases}\label{eq:sens-diff-summary}
\end{align}

At $N\sim 1$ this sensitivity is on order of the sensitivity of classical speckles, but it increases dramatically for $N\gg 1$. The qualitative explanation for this criterion is as follows: A classical wave propagates through the sample via a series of channels, undergoing $(L/\ell)^2$ scattering events per channel. Changing an external parameter $\gamma$ will change the phase accumulated at each scattering event by some quantity $\phi_0(\gamma)$, and the total phase change will also accumulate diffusively so that the total change of phase per channel is $\sim\phi_0(\gamma) L/\ell $. The speckle pattern will change significantly when this phase difference is of order one. In the case of directed waves, the phase difference equals $(\bar{k}L\theta_L)\phi$ for a change in incidence angle $\phi$, and $(D_\theta L^2/c)\Delta \omega  \sim (\theta_L^2 L/c)\Delta \omega $ for a frequency change. The sensitivity of an $N$-photon speckle is to leading order given by a product of the sensitivities of each detector measurement, and so we expect a further strong dependence of the sensitivity on $N$.

As an illustrative example let us focus on a change of incidence angle in the directed wave regime. 
We assume that it is sufficiently small that its effect on the phase of a trajectory going from one end of the sample to the other is small, namely
\be
\phi < 1/(\theta_L \bar{k} L) \label{condition}.
 \ee
Evaluating the diagrams in Fig. \ref{fig8} we find
\begin{widetext}
\be
\label{eq:sens-directed}
\mathcal K(\phi;N)=\langle I_N \rangle^{2}\sum _{j=0}^{N-1} \left( \begin{array}{c} N \\ j \end{array} \right) \mbox{Tr} \left( \rho^{(N-j)}\right)^2 \exp\left[ -\frac{N-j}{3} (\theta_L \bar{k} L \phi)^2 \right].
\ee
For pure non-entangled states $\mbox{Tr} \left( \rho^{(\nu-j)}\right)^2=1$ for all $j$ and we obtain:
\be
\frac{\mathcal K(\phi;N)}{\mathcal K(0;N)}= \frac{ \left( 1+ e^{-\frac{1}{3} (\theta_L \bar{k} L \phi)^2} \right)^N-1}{2^N-1} \xrightarrow[N\gg1]{} e^{- \frac{N}{6} (\theta_L \bar{k} L \phi)^2} \label{eq:limit}
\ee
\end{widetext}
In the directed wave regime, the classical correlation function decays as a Gaussian, as shown in Eq. \eqref{eq:sens-directed}. As a result the typical angle $\phi_N^*$ is proportional to $1/\sqrt{N}$. 

In the diffusive regime, the decay of the classical correlation function associated with short range correlations is exponential, $\mathcal{K}(\phi;1) \sim \exp(-\phi/\phi^*)$ where $\phi^* = 1/\bar{k}L$ (see Eq. \eqref{eq:C0-diffusive}). Similar considerations to those used in the directed wave regime imply that $\phi_N^*$ is inversely proportional to $N$, yielding Eq. \eqref{eq:sens-diff-summary}.

To discuss the sensitivity to the change in the scattering potential, we have to introduce parameters describing these changes. In the diffusive case, it is customary to characterize the change by the number of impurities shifted from their initial positions. It is assumed that the amplitude of the scattering length, and the spatial shifts, are both  on order of the wavelength. A repetition of the qualitative arguments mentioned above yields a typical number of impurities 
\begin{equation}
n_N^* =  \mathcal{A} \bar{k}^2 \ell / L N.
\end{equation}
In the directed wave regime, one cannot treat the disorder as a series of strong (S-wave) scatterers. Thus, the sensitivity should depend on the precise form of the change in the scattering potential. Qualitatively we expect the behaviour to be the same as for the other parameter changes we analyzed, i.e. the sensitivity will depend in some exponential way on both the strength of the change and the number of photons. However, we shall not pursue the subject further in this work.

\section{Summary}

In this paper we studied the statistics of speckle patterns of $N$-entangled photons propagating in disordered systems. Most previous studies in this field concentrated on the Random Matrix Theory regime. Our study moves beyond this, focusing on corrections beyond RMT, associated with long-range correlations caused by diffraction.

We showed the existence of long range correlations in these speckle patterns. To a leading order approximation in $1/\bar{k}\ell$ these depend only on the reduced density matrix  Tr$\left(\rho^{(1)}\right)^2$, namely they are a single-photon property. The reason behind this can be traced to a simple phase space argument. Multiphoton interference effects depend on coincident crossings of classical trajectories in the bulk. The phase space for each such crossing is inversely proportional to the cross-sectional area, and therefore such multiple crossings are strongly suppressed. 

Nevertheless, it turns out that the long range correlations of $N$-photon speckle patterns determine the results measured by large aperture detectors. 

We also showed that the sensitivity of the $N$-photon speckle pattern to change of parameters is greatly enhanced. This enhancement is by factor of $N$ for the case of real space diffusion, and by a factor of $\sqrt{N}$ for directed waves.      
   
We wish to point out that our analysis is relevant to much broader problems than the specific system of entangled photons in quenched disorder described above. For example, many of our results apply to problems of diffusion in systems of coupled quantum bits. Random benchmarking of quantum computation systems \cite{Emerson2005,Knill2008,Barends2014}, is rapidly emerging as a powerful diagnostic tool and our work provides a further framework for measurement and analysis in this field.

\acknowledgments
We wish to thank Y. Bromberg, N. Katz and O. Rosolio for stimulating discussions. This research was supported by the United States-Israel Binational Science Foundation (BSF) grant No. 2012-134, and the  Israel Science Foundation (ISF) grant No, 302/14.

\newpage\appendix
\onecolumngrid
\section{Derivation of the correlation function ${\cal C}(\bm{k}_1, \bm{k}_1')$}
\label{sec:app-two-point-corr}
In this appendix we show that the leading order diagrams for $\mathcal{K}(\hbk_1,\hbk_2,\hbk_1',\hbk_2')$ are those shown in Fig. \ref{fig6}, and calculate the correlation function \eqref{eq:2-photon-corr}.

Consider the general expression for the correlation function,
\begin{flalign}
  \label{eq:corr-general}
  \mathcal{K}(\hbk_1,\hbk_2,\hbk_1',\hbk_2') = &\beta_N^2 \sum_{q_1,q_2,\tilde{q}_1,\tilde{q}_2} \sum_{q_1',q_2',\tilde{q}_1',\tilde{q}_2'}\rho_{q_1q_2;\tilde{q}_1\tilde{q}_2}\rho_{q_1'q_2';\tilde{q}_1'\tilde{q}_2'}\\ &\langle\delta(\psi_{\bm{q}_1}(\k_1)\psi_{\bm{q}_2}(\k_2)\psi^*_{\tilde{\bm{q}}_1}(\k_1)\psi_{\tilde{\bm{q}}_2}(\k_2))
\delta(\psi_{\bm{q}_1'}(\k_1')\psi_{\bm{q}_2'}(\k_2')\psi^*_{\tilde{\bm{q}}_1'}(\k_1')\psi_{\tilde{\bm{q}}_2'}(\k_2'))\rangle.\nonumber
\end{flalign}
Most of the terms in the above sum can be neglected. The terms yielding the largest contribution are those appearing in Fig.~\ref{fig6}a,c. For example the first diagram on the left of Fig. \ref{fig6}a is obtained by constraining the sum to terms with
\begin{equation*}
  \label{eq:6a-left-matching}
  q_1 = \tilde q_1',~~q_1' = \tilde q_2,~~q_2 = \tilde q_2,~~~q_2'=\tilde q_2'.
\end{equation*}
In what follows we evaluate the diagrams of Fig. \ref{fig6}, and show that they constitute the leading contribution. We do so using the Langevin scheme for the directed-wave regime\cite{Agam2007}, which we briefly review here.

First, let us introduce a ray distribution function, given as a Wigner transform of the product of retarded Green function and advanced Green function: 
\begin{eqnarray}
  f({\bf R}-{\bf R}',\bm k-\bm k'; z) = \int d^2 \delta r ~ d^2 \delta r'~ G\left( {\bf R}+\frac{{\bf \delta r}}{2},0;{\bf R}'-\frac{{\bf \delta r}'}{2},z\right) G^*\left( {\bf R}-\frac{{\bf \delta r}}{2},0;{\bf R'}+\frac{{\bf \delta r'}}{2},z\right) e^{-i\bar{k} \left( \bm k\cdot{\bf \delta r}+ \bm k'\cdot{\bf \delta r}'\right)} \label{Av1}.
\end{eqnarray}
We decompose the distribution function $f(\bm R, \bm k, z)$ into an average and fluctuating part,
\begin{equation}
  \label{eq:f-decomposition}
  f({\bf R}, {\bf k};z)= \langle f({\bf R}, {\bf k};z) \rangle + \delta f({\bf R}, {\bf k};z).
\end{equation}
The function $\langle f(\bm R, \bm k, z)\rangle$ satisfies the Boltzmann equation
\be
\frac{\partial\langle f({\bf R},\bm k;z) \rangle}{\partial z} + \hbk \cdot\frac{\partial\langle f({\bf R},\bm k;z) \rangle}{\partial {\bf R}}=I_{st}[\langle f({\bf R},\bm k;z) \rangle]\equiv \int d^2 k' {\cal G}(\bm k-\bm k')\left[\langle f({\bf R},\bm k';z) \rangle-\langle f({\bf R},\bm k;z) \rangle\right] \label{Boltzmann}
\ee
while the fluctuating part obeys the the Langevin  equation,
\begin{equation}
  \label{eq:df-eqs-1}
  \frac{\partial\delta  f({\bf R},{\bf k};z)}{\partial z} + {\bf k} \cdot \frac{\partial\delta  f({\bf R},{\bf k};z)}{\partial {\bf R}}
=I_{st}[\langle f({\bf R},\bm k;z) \rangle] + \mathcal{L}({\bf R},{\bf k};z),
\end{equation}
where the Langevin sources have zero mean and correlation function,
\begin{align}
\langle \mathcal{L}({\bf R},\bm k;z) \mathcal{L}({\bf R}',\bm k';z')\rangle &= \frac{2 \pi}{\bar{k}^2}\delta_{\alpha\beta}\delta(\bm R - \bm R')\delta(z-z')\times \nonumber \\
&\left[\delta(\bm k - \bm k') f_+(\bm R,\bm k,z)\int d^2\tilde{k}~\mathcal{G}(\bm k-\tilde{\bm k}) f_-(\bm R, \tilde{\bm k},z) -  f_+(\bm R,\bm k,z) \mathcal{G}(\bm k-\bm k') f_-(\bm R, \bm k',z')\right]  \label{eq:df-eqs-2},
\end{align}
and $f_\pm$ obey Eq. \eqref{Boltzmann} as well, with boundary conditions
\begin{equation}
\label{eq:directed-langevin-source-2}
f_+({\bf R},{\bf u };0) =  |A_0|^2 \delta ({\bf u}-{\bf q}),~~~~~~~~
f_-({\bf R},{\bf u };0) =   |A_0|^2 \delta ({\bf u}-{\bf q}'),
\end{equation}
for incoming plane waves in directions $\bm q,\bm q'$.

On length scales much larger than the elastic mean free path $\ell$, and angles much larger than $\theta_0$ the Boltzmann equation (\ref{Boltzmann}) reduces to the diffusion-like equation:
 \be
\frac{\partial\langle f({\bf R},\bm k;z) \rangle}{\partial z} + \bm k \cdot\frac{\partial\langle f({\bf R},\bm k;z) \rangle}{\partial {\bf R}}-
 D_\theta \frac{\partial^2 \langle f({\bf R},\bm k;z)\rangle}{\partial \hbk^2}=0,
\ee
where $D_\theta = 1/2\ell_{tr}$ is the diffusion constant in the angle space. The solution of this equation for boundary conditions $\langle f({\bf R},\bm k;0) \rangle= \delta({\bf R}) \delta(\bm k)$ is
\begin{eqnarray}
\langle  f({\bf R},\bm k;z)\rangle=\frac{3}{4\pi^2 D_\theta^2 z^4} \exp \left[ -\frac{ 3{\bf R}^2}{D_\theta z^3}+ \frac{3 \bm k \cdot {\bf R}}{D_\theta z^2}- \frac{ \bm k^2}{D_\theta z} \right].
\end{eqnarray}

Furtheremore in this diffusive regime Eqs. \eqref{eq:df-eqs-1}-\eqref{eq:df-eqs-2} simplify to
\begin{align}
  \mathcal{L}(\bm R,\bm k;z) &= \nabla_{\bm k} \cdot \bm j(\bm R,\bm k;z) \\
\langle j_\alpha^L({\bf R},\bm k;z) j_\beta^L({\bf R}',\bm k';z')\rangle &= \frac{2 \pi D_\theta}{\bar{k}^2} f_+({\bf R},\bm k;z)  f_-({\bf R}',\bm k';z)   \delta_{\alpha\beta} \delta({\bf R}-{\bf R}') \delta(\bm k-\bm k')\delta(z-z') \label{eq:JJcorrelator}
\end{align}
where $f_\pm({\bf R},{\bf u};z)$ satisfy the equation:
\be
\frac{\partial f_\pm({\bf R},{\bf u};z)}{\partial z} +{\bf u} \cdot \frac{\partial  f_\pm({\bf R},{\bf u};z)}{\partial {\bf R}}
- D_\theta \frac{\pd^2 f_\pm ({\bf R},{\bf u};z)}{\pd \bm{u}^2}=0 \label{eq:fpm}
\ee
with boundary conditions \eqref{eq:directed-langevin-source-2}.

The correlation function we seek to calculate is
\begin{equation}
\label{Ckk-def}
\mathcal{C} ({\bf k}, {\bf k}') =\frac{1}{I_1^2} \int d^2R d^2R' \langle \delta f({\bf R}, {\bf k};L) \delta f({\bf R}', {\bf k}';L)\rangle
\end{equation}
where
\be
I_1=  \int d^2R \langle f({\bf R}, {\bf k};L)\rangle = \frac{|A_0|^2 {\cal A}}{2\pi \theta_L^2}
\ee
(assuming $q_1,q_1',k \ll \theta_L$.)

\subsection{Long range correlations $\theta_0 \ll|\hbk_i-\hbk_j|$}
\label{sec:long-range-corr}

We begin by finding the long-range correlations, when $|\bm k_i - \bm k_j| \gg \theta_0$, i.e. when the rays undergo many scattering events after crossing. Solving Eq.~(\ref{eq:fpm}) and substituting the result in  Eq.~(\ref{eq:JJcorrelator}) we obtain
\begin{eqnarray}
\langle j_\alpha^L({\bf R}_{1},{\bf u};z) j_\beta^L({\bf R}_{1}',{\bf u}';z')\rangle &=& \frac{ |A_0|^4 }{8 D_\theta  \pi \bar{k}^2 z^2}  \exp\left[  -\frac{({\bf u} - {\bf q})^2+({\bf u}' - {\bf q}')^2 }{4 D_\theta z} \right] \delta_{\alpha\beta} \delta({\bf R}_1-{\bf R'}_1) \delta({\bf u}-{\bf u}')\delta(z-z')
\end{eqnarray}

Thus
\begin{eqnarray}
&&\langle \delta f({\bf R}, {\bf k};L) \delta f({\bf R}', {\bf k}';L)\rangle = \int d^2 R_1 dz d^2 R'_1 dz' d^2 u d^2 u'
\langle f({\bf R}-{\bf R}_1; {\bf k}- {\bf u}; L-z)\rangle \langle f({\bf R}'-{\bf R}'_1; {\bf k}'-{\bf u}'; L-z')\rangle \nonumber \\
&&\times \frac{ \partial}{\partial u_{\alpha }} \frac{ \partial}{\partial u'_{\beta} }\frac{|A_0|^4}{8 D_\theta  \pi \bar{k}^2 z^2}  \exp\left[  -\frac{({\bf u} - {\bf q})^2+({\bf u}' - {\bf q}')^2}{4 D_\theta z} \right] \delta_{\alpha\beta} \delta({\bf R}_1-{\bf R}_1') \delta({\bf u}-{\bf u}') \delta(z-z')
\end{eqnarray}
Integration by parts gives
\begin{eqnarray}
\langle \delta f({\bf R}, {\bf k};L) \delta f({\bf R}', {\bf k}';L)\rangle &=&
 \int d^2 R_1 d^2 u dz
\frac{ \partial}{\partial{\bf k}}\langle f({\bf R}-{\bf R}_1; {\bf k}-{\bf u};L-z)\rangle \cdot \frac{ \partial}{\partial {\bf k}'}\langle f({\bf R}'-{\bf R}_1; {\bf k}'-{\bf u}; L-z) \rangle \nonumber \\
&\times & \frac{ |A_0|^4 }{8 D_\theta  \pi \bar{k}^2 z^2}  \exp\left[  -\frac{({\bf u} -{\bf q})^2+({\bf u} -{\bf q}')^2}{4 D_\theta z}  \right]. \nonumber
\end{eqnarray}
Plugging this into Eq. \eqref{Ckk-def} yields
\begin{eqnarray}
\mathcal{C} ({\bf k}, {\bf k}') &=& \frac{1}{I_1^2}\int d z d^2u\frac{ \mathcal{A}}{(4\pi)^2 D_\theta^2 (L-z)^2}\frac{ \partial}{\partial{\bf k}}\cdot \frac{ \partial}{\partial{\bf k}'} \exp \left[- \frac{ ({\bf k}-{\bf u})^2+({\bf k}'-{\bf u})^2 }{4 D_\theta (L-z)} \right] \nonumber \\
&\times &\frac{ |A_0|^4 }{8 D_\theta  \pi \bar{k}^2 z^2}  \exp\left[ -\frac{({\bf u} -{\bf q})^2+({\bf u} -{\bf q}')^2}{4 D_\theta z} \right] \nonumber 
\end{eqnarray}
Finally, changing variables to $\zeta= z/L$ and performing the $d^2u$ integral, we arrive at
\begin{equation}
\label{eq:directed-C}
\mathcal{C}({\bf k},{\bf k}')= \frac{1 }{4 \bar{k}^2 \mathcal{A}} \int_{0}^{1} \frac{d \zeta}{(1-\zeta)\zeta}\frac{\partial}{\partial{\bf k}}\cdot \frac{ \partial}{\partial{\bf k}'}  \exp \left[ -\frac{({\bf k}-{\bf k}')^2}{4\theta_L^2 (1-\zeta)} -\frac{({\bf k}+ {\bf k}'- {\bf q}-{\bf q}')^2}{4 \theta_L^2}-\frac{({\bf q}-{\bf q}')^2}{4 \theta_L^2 \zeta}\right].
\end{equation}
\begin{wrapfigure}{r}{0.4\textwidth}
  \centering
  \includegraphics[width=0.2\textwidth]{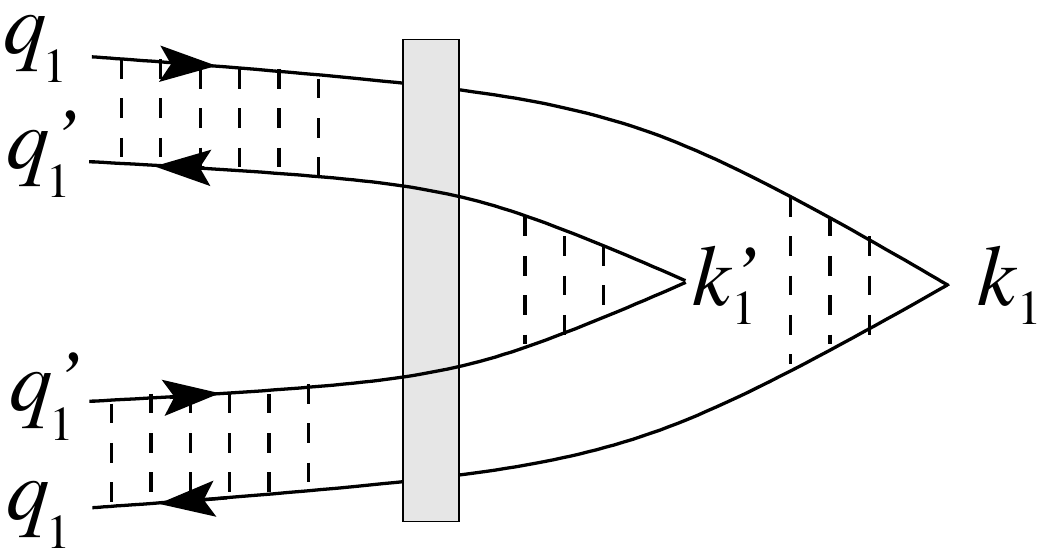}
  \caption{A subleading diagram contributing to ${\cal C}(\bm{k}_1, \bm{k}_2)$.}
  \label{fig:diag-appendix}
  \vspace{0.3cm}
\end{wrapfigure}
Performing the integral and specializing to $\theta_L\gg|\hbk- \hbk'|\gg \theta_0$ we end up with the second line of Eq. \eqref{C0}.

Next we show that the contribution we just calculated is the dominant one. To do so it is enough to evaluate the diagram in Fig. \ref{fig:diag-appendix}. The boundary conditions in this case are
\be
f_\pm({\bf R},\bm{q};0) = |A_0|^2 \exp ( \pm i \bar{k} \Delta \bm{q} \cdot {\bf R} ) \delta (\bm{q}-\bar{\bf q})
\ee
where  $\Delta \bm{q}=\bm{q}_1-\bm{q}_1'$, and  $\bar{\bm q}=(\bm{q}_1+\bm{q}_1')/2$. Furthermore we assume $\bar{k}L\theta_L \Delta\bm{q} \gg 1$. Repeating the steps above and comparing to expression \eqref{C0}, we find the result to be smaller by an order of $\frac{\ell} {L}$.

\subsection{Intermediate range correlations: $\theta_W \ll|\hbk_i-\hbk_j| \ll \theta_0$}
\label{sec:Intermediate-range-corr}

The integral (\ref{eq:directed-C}) diverges logarithmicaly when  $|\hbk_i-\hbk_j| \ll \theta_0$. This is because of the breakdown of the diffusive approximation, as for small angles the main contribution comes from the last scattering event before the ray leaves the sample. Thus when $\theta_W \ll |\hbk_i-\hbk_j| \ll \theta_0$, we can neglect the diffusion term in Eq. \eqref{eq:df-eqs-1}, and simply solve for the correlation function to first order in the scattering probability $\mathcal G$. In this limit Eq. \eqref{eq:df-eqs-1} can be easily solved and we find
\begin{equation}
  \label{eq:short-df}
  \delta f(\bm R, \hbk, z) = \int^z d\zeta \mathcal{L}(\bm R + (\zeta - z) \hbk, \hbk, \zeta).
\end{equation}
We then evaluate Eq.~\eqref{Ckk-def} directly (to leading order in $\mathcal G$) and find
\begin{equation}
  \label{eq:C0-mid}
  \mathcal{C}(\hbk_1,\hbk_2) = \frac{2\pi}{\bar{k}^2\mathcal{A}}\left[\delta(\hbk_1-\hbk_2) - \ell \mathcal{G}(|\hbk_1-\hbk_2|)\right].
\end{equation}

The first term can be understood as a small correction to Eq. \eqref{C-Short} arising from short range interference that occurs when two rays meet. The second term comes from impurity scattering and represents the fact that in the directed-wave regime, all ray meetings occur when rays have both the same position and the same direction of propagation. This is different from the diffusive case, where many rays moving in different directions may cross at the same point. Thus, in the directed-wave regime the free propagation of the rays from a given point implies that they cannot scatter to an angle $\sim\theta_0$, giving rise to the negative correlation. Neglecting the first term in Eq. \eqref{eq:C0-mid} we find the first line in Eq. \eqref{C0}.


\subsection{Short range correlations: $|\hbk_i-\hbk_j| \ll \theta_W$}
\label{sec:short-range-corr}
When $|\hbk_i-\hbk_j| \ll \theta_W$, one can neglect the correlations from ray diffraction (Hikami Box). In this limit we may treat $\psi_\bm{q}(\k)$ as independent Gaussian variables satisfying the relation
\begin{align}
  \label{eq:diff-fourier}
 \alpha(\k,\k')= \langle \psi_\bm{q}(\k)\psi^*_\bm{q}(\k')\rangle  
= \int d^2R \langle f(\bm R, \frac{\hbk + \hbk'}{2}, L) \rangle e^{i\bar{k}(\hbk-\hbk')\cdot\bm R}.
\end{align}
and the correlation function we seek to calculate is ${\cal C}(\k,\k')=|\alpha(\k,\k')/\alpha(\k,\k)|^2$.
Assuming $|\hbk|,|\hbk'|\ll \theta_L$ and that $\langle f(\bm R, \hbk, L) \rangle$ is independent of $\bm R$ throughout the slab (which is the case when $W \gg \theta_L L$) we obtain Eq. \eqref{C-Short}.

\section{Derivation of the multiphoton correlation function in the diffusive regime}

\label{sec:app-diffusive-regime}

In this appendix we sketch out the steps leading to eqs. \eqref{eq:mainD}. Just as in the directed wave regime, the leading order corrections are given by the diagrams of Fig. \ref{fig6}. However, in this regime we have real space diffusion, and the distribution function $f(\bf R,\bm{q}; z)$ is replaced by the local intensity, obtained by integrating out the angular part of $f$,
\be
I({\bf R}; z) = \langle I({\bf R}; z)\rangle + \delta I({\bf R}; z),
\ee
where $\delta I$ is the fluctuating part of the intensity. Referring to Eq. \eqref{eq:corr-general} we see that we must evaluate a product of four Green's functions, as in the directed-wave case, e.g.
\begin{equation}
  \label{eq:4-green-correlator}
  \langle \psi_\bm{q}(\k)\psi^*_{\tilde{\bm{q}}}(\k) \psi_{\bm{q}'}(\k')\psi^*_{\tilde{\bm{q}}'}(\k')\rangle = \int e^{i (\bm{q} \cdot \bm{r} - \k \cdot \bm{s})}e^{-i (\tilde{\bm{q}} \cdot \tilde{\bm{r}} - \k \cdot \tilde{\bm{s}})}e^{i (\bm{q}' \cdot \bm{r}' - \k' \cdot \bm{s}')}e^{-i (\tilde{\bm{q}}' \cdot \tilde{\bm{r}}' - \k' \cdot \tilde{\bm{s}}')}G(\bm{r}, \bm{s})G^*(\tilde{\bm{r}}, \tilde{\bm{s}})G(\bm{r}', \bm{s}')G^*(\tilde{\bm{r}}', \tilde{\bm{s}}').
\end{equation}
where we have suppressed the explicit $L$ dependence of the Green's functions. It is clear that upon disorder averaging there are two ways to match Green's functions into Diffusons, namely:
\begin{align}
  \label{eq:diff-matching}
  \bm{r} = \tilde{\bm{r}}', \bm{r}' = \tilde{\bm{r}}, \bm{s} = \tilde{\bm{s}}, \bm{s}' = \tilde{\bm{s}}' &\Rightarrow \bm{q} = \tilde{\bm{q}}', \bm{q}' = \tilde{\bm{q}} \\
  \bm{r} = \tilde{\bm{r}}, \bm{r}' = \tilde{\bm{r}}', \bm{s}' = \tilde{\bm{s}}, \bm{s} = \tilde{\bm{s}}' &\Rightarrow \bm{q} = \tilde{\bm{q}}, \bm{q}' = \tilde{\bm{q}}'
\end{align}
These two matchings are depicted in Fig. \ref{fig:diffusive-hikami}. The constraints on the incoming $q$'s can be obtained straightforwardly by evaluating the two diagrams using well known results\cite{Akkermans2007}. The result is:
\begin{equation}
  \label{eq:hikami-diffuse-final}
  \mathcal{C}_{q,\tilde{q},q',\tilde{q}'}(\k,\k') = \delta_{q\tilde{q}'}\delta_{q'\tilde{q}} \tilde{\mathcal C}(0) + \delta_{q\tilde{q}}\delta_{q'\tilde{q}'}\tilde{\mathcal C}(|\hat{\k}-\hat{\k'|}),
\end{equation}
where $\tilde{\mathcal C}(k)$ was described in eq. \eqref{eq:C0-diffusive}. Generalizing to $N$ photons and performing the necessary summation we get our final result, Eq. \eqref{eq:mainD1}.

\twocolumngrid
\bibliographystyle{unsrt}
\bibliography{speckles}

\end{document}